\documentclass[12pt]{article}
\usepackage{graphicx}

\usepackage{epsfig}
\usepackage{url}
\usepackage[centertags]{amsmath}
\usepackage{amssymb,amscd}
\usepackage{hyperref}

\def\beq{\begin{equation}}
\def\eeq{\end{equation}}

\usepackage{url,hyperref,lineno,microtype,subcaption}

\newcommand\pubnumber{SNSN-323-63}
\newcommand\pubdate{\today}

\def\address{Wisconsin IceCube Particle Astrophysics Center\\
Madison, WI USA}

\def\Title#1{\begin{center} {\Large #1 } \end{center}}
\def\Author#1{\begin{center}{ \sc #1} \end{center}}
\def\Address#1{\begin{center}{ \it #1} \end{center}}

\newcommand\pubblock{\rightline{\begin{tabular}{l} \pubnumber\\
         \pubdate  \end{tabular}}}
\newenvironment{Abstract}{\begin{quotation}  }{\end{quotation}}

\def\beq{\begin{equation}}
\def\eeq#1{\label{#1}\end{equation}}
\def\eeqn{\end{equation}}

\def\beqa{\begin{eqnarray}}
\def\eeqa#1{\label{#1}\end{eqnarray}}
\def\eeqan{\end{eqnarray}}

\let\bar=\overbar

\def\Dslash{\not{\hbox{\kern-4pt $D$}}}
\def\dslash{\not{\hbox{\kern-2pt $\del$}}}

\def\msb{{\bar{\ssstyle M \kern -1pt S}}}

\begin{document}
\begin{titlepage}
\pubblock

\vfill
\Title{The Anisotropies and Origins of Ultrahigh Energy Cosmic Rays}
\vfill
\Author{Francis Halzen}
\Address{\address}
\vfill
\begin{Abstract}
After updating the status of the measurements of the cosmic neutrino flux by the IceCube experiment, we summarize the observations of the first identified source of cosmic rays and speculate on the connection between the two observations.
\end{Abstract}
\vfill
\vfill
\end{titlepage}
\def\thefootnote{\fnsymbol{footnote}}
\setcounter{footnote}{0}

\section{Detecting cosmic neutrinos}\label{sec1}

Cosmic rays have been studied for more than a century. They reach energies in excess of $10^8$\,TeV, populating an extreme universe that is opaque to electromagnetic radiation~\cite{Kotera:2011cp,Ahlers:2015lln}. We don't yet know where or how particles are accelerated to these extreme energies, but with the observation of a distant blazar in coincidence with the direction and time of a very high energy muon neutrino, neutrino astronomy has made a breakthrough in resolving this puzzle \cite{IceCube:2018dnn,IceCube:2018cha}. The rationale for finding cosmic ray sources by observing neutrinos is straightforward: near neutron stars and black holes, gravitational energy released in the accretion of matter can power the acceleration of protons ($p$) or heavier nuclei that subsequently interact with gas (``$pp$'') or ambient radiation (``$p\gamma$'') to produce neutrinos originating from the decay of pions and other secondary particles. In the case of photoproduction, both neutral and charged pion secondaries are produced in the processes $p+\gamma_{\rm bg}\to p+\pi^0$ and $p+\gamma_{\rm bg}\to n+\pi^+$, for instance. While neutral pions decay as $\pi^0\to\gamma+\gamma$ and create a flux of high-energy gamma rays, the charged pions decay into three high-energy neutrinos ($\nu$) and anti-neutrinos ($\bar\nu$) via the decay chain $\pi^+\to\mu^++\nu_\mu$ followed by $\mu^+\to e^++\bar\nu_\mu +\nu_e$, and the charged-conjugate process. We refer to these photons as pionic photons. They provide the rational for multimessenger astronomy and should be distinguished from photons radiated by electrons that may be accelerated along with the cosmic rays.

\begin{figure}[t]
  \centering
    \includegraphics[width=1.0\linewidth]{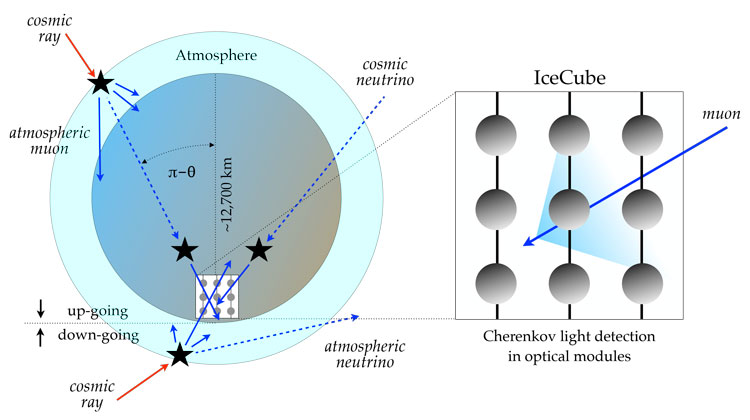}
  \caption{The principal idea of neutrino telescopes from the point of view of IceCube located at the South Pole. Neutrinos dominantly interact with a nucleus in a transparent medium like water or ice and produce a muon that is detected by the wake of Cherenkov photons it leaves inside the detector. The background of high-energy muons (solid blue arrows) produced in the atmosphere can be reduced by placing the detector underground. The surviving fraction of muons is further reduced by looking for upgoing muon tracks that originate from muon neutrinos (dashed blue arrows) interacting close to the detector. This still leaves the contribution of muons generated by atmospheric muon neutrino interactions. This contribution can be separated from the diffuse cosmic neutrino emission by an analysis of the combined neutrino spectrum.}
    \label{fig:earth}
\end{figure}
 
High-energy neutrinos interact predominantly with matter via deep inelastic scattering off nucleons: the neutrino scatters off quarks in the target nucleus by the exchange of a $Z$ or $W$ weak boson, referred to as {\it neutral current} and {\it charged current} interactions, respectively. Whereas the neutral current interaction leaves the neutrino state intact, in a charged current interaction a charged lepton is produced that shares the initial neutrino flavor. The average relative energy fraction transferred from the neutrino to the lepton is at the level of $80$\% at high energies. The struck nucleus does not remain intact and its high-energy fragments typically initiate hadronic showers in the target medium.

\begin{figure}[t]
  \centering
   \includegraphics[width=1.0\linewidth]{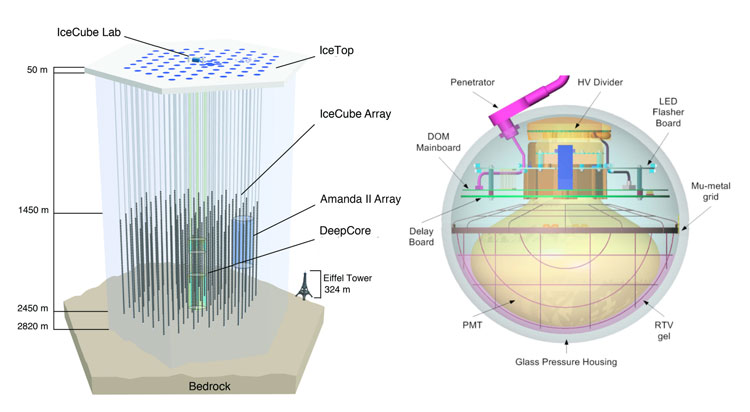}
  \caption[]{Architecture of the IceCube observatory (left) and the schematics of a digital optical module (right) (see Ref.~\cite{Abbasi:2008aa} for details).}
  \label{fig:deepcore}
\end{figure}

Immense particle detectors are required to collect cosmic neutrinos in statistically significant numbers. Already by the 1970s, it had been understood~\cite{Roberts:1992re} that a kilometer-scale detector was needed to observe the cosmogenic neutrinos produced in the interactions of cosmic rays with background microwave photons~\cite{Beresinsky:1969qj}. The IceCube project has transformed one cubic kilometer of natural Antarctic ice into a Cherenkov detector. Photomultipliers embedded in the ice transform the Cherenkov light radiated by secondary particles produced in neutrino interactions into electrical signals using the photoelectric effect; see Figs.~\ref{fig:earth} and \ref{fig:deepcore}. Computers at the surface use this information to reconstruct the light patterns produced to infer the arrival directions, energies and flavor of the neutrinos.

Two patterns of Cherenkov radiation are of special interest, ``tracks'' and ``cascades.'' The term ``track'' refers to the Cherenkov emission of a long-lived muon passing through the detector after production in a charged current interaction of a muon neutrino inside or in the vicinity of the detector. Because of the large background of muons produced by cosmic ray interactions in the atmosphere, the observation of muon neutrinos is limited to upgoing muon tracks that are produced by neutrinos that have passed through the Earth that acts as a neutrino filter. The remaining background consists of atmospheric neutrinos, which are indistinguishable from cosmic neutrinos on an event-by-event basis. However, the steeply falling spectrum ($\propto E^{-3.7}$) of atmospheric neutrinos allows identifying diffuse astrophysical neutrino emission above a few hundred TeV by a spectral analysis. The atmospheric background is also reduced for muon neutrino observation from point-like sources, in particular transient neutrino sources.

Energetic electrons and taus produced in interactions of electron and tau neutrinos, respectively, will initiate an electromagnetic shower that develops over less than 10 meters. This shower as well as the hadronic particle shower generated by the target struck by a neutrino in the ice  radiate Cherenkov photons. The light pattern is mostly spherical and referred to as a  ``cascade.'' The direction of the initial neutrino can only be reconstructed from the Cherenkov emission of secondary particles produced close to the neutrino interaction point, and the angular resolution is worse than for track events. On the other hand, the energy of the initial neutrino can be constructed with a better resolution than for tracks.

\begin{figure}[t]\centering
\includegraphics[width=0.43\linewidth]{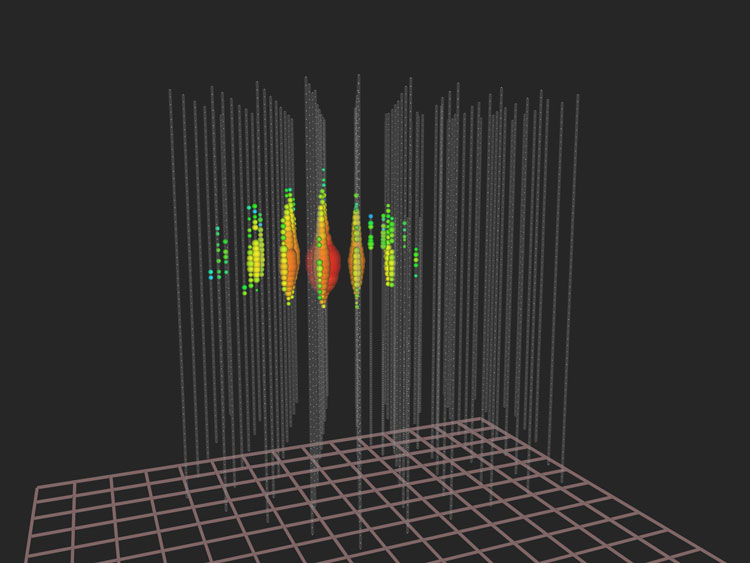}\hspace{0.5cm}\includegraphics[width=0.43\linewidth]{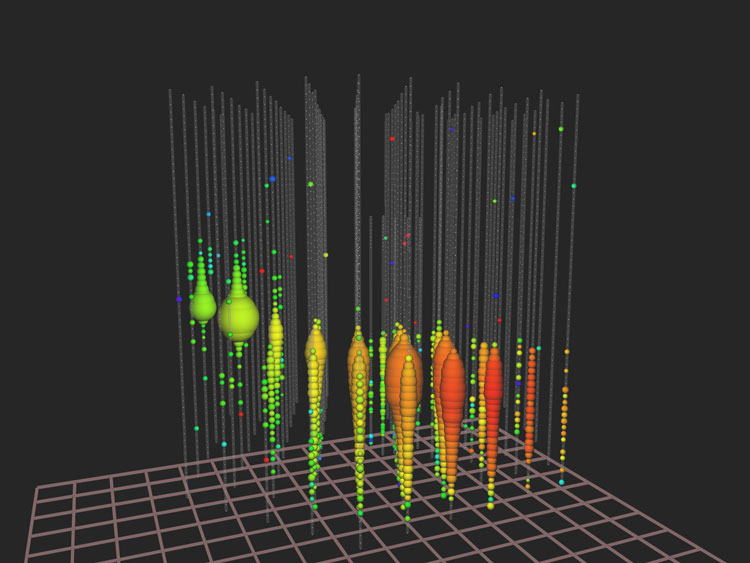}
\caption[]{{\bf Left Panel:}  Light pool produced in IceCube by a shower initiated by an electron or tau neutrino. The measured energy is $1.14$ PeV, which represents a lower limit on the energy of the neutrino that initiated the shower. White dots represent sensors with no signal. For the colored dots, the color indicates arrival time, from red (early) to purple (late) following the rainbow, and size reflects the number of photons detected. {\bf Right Panel:}  An upgoing muon track traverses the detector at an angle of $11^\circ$ below the horizon. The deposited energy, i.e., the energy equivalent of the total Cherenkov light of all charged secondary particles inside the detector, is 2.6\,PeV.}
\label{fig:erniekloppo}
\end{figure}

Two methods are used to identify cosmic neutrinos. Traditionally, neutrino searches have focused on the observation of muon neutrinos that interact primarily outside the detector to produce kilometer-long muon tracks passing through the instrumented volume. Although this allows the identification of neutrinos that interact outside the detector, it is necessary to use the Earth as a filter in order to remove the background of cosmic-ray muons. This limits the neutrino view to a single flavor and half the sky. An alternative method exclusively identifies high-energy neutrinos interacting inside the detector, so-called high-energy starting events (HESE). It divides the instrumented volume of ice into an outer veto shield and a $\sim420$-megaton inner fiducial volume. The advantage of focusing on neutrinos interacting inside the instrumented volume of ice is that the detector functions as a total absorption calorimeter, measuring the neutrino energy of cascades with a 10-15\,\% resolution~\cite{Aartsen:2013vja}. Furthermore, with this method, neutrinos from all directions in the sky can be identified, including both muon tracks as well as secondary showers, produced by charged-current interactions of electron and tau neutrinos, and neutral current interactions of neutrinos of all flavors. For illustration, the Cherenkov patterns initiated by an electron (or tau) neutrino of about~1\,PeV energy and a muon neutrino losing 2.6\,PeV energy in the form of Cherenkov photons while traversing the detector are contrasted in Fig.~\ref{fig:erniekloppo}.

In general, the arrival times of photons at the optical sensors, whose positions are known, determine the particle's trajectory, while the number of photons is a proxy for the deposited energy, i.e., the energy of all charged secondary particles from the interactions. For instance, for the cascade event shown in the left panel of Fig.~\ref{fig:erniekloppo}, more than 300 digital optical modules (DOMs) report a total of more than 100,000 photoelectrons. The two abovementioned methods of separating neutrinos from the cosmic-ray muon background have complementary advantages. The long tracks produced by muon neutrinos can be pointed back to their sources with a $\le 0.4^\circ$ angular resolution.  In contrast, the reconstruction of the direction of cascades in the HESE analysis, in principle possible to a few degrees, is still in the development stage in IceCube~\cite{Aartsen:2013vja}. Their reconstruction, originally limited to within $10^\circ\sim15^\circ$ of the direction of the incident neutrino, has now achieved resolutions closer to $5^\circ$ \cite{Tyuan2017}.
Determining the deposited energy from the observed light pool is, however, relatively straightforward, and a resolution of better than 15\,\% is possible; the same value holds for the reconstruction of the energy deposited by a muon track inside the detector.

\begin{figure}[ht!]\centering
\includegraphics[width=0.7\linewidth]{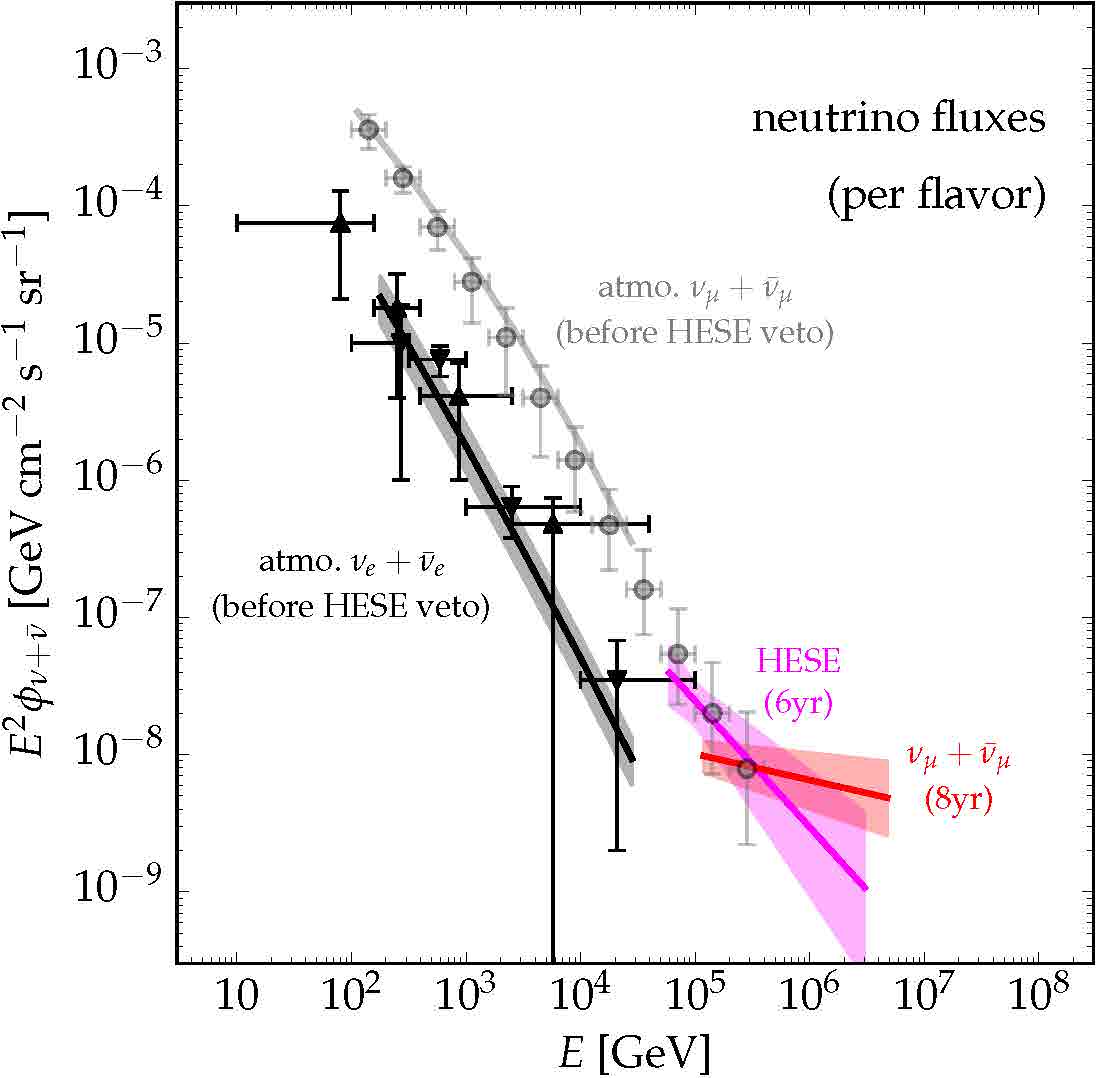}
\caption[]{Summary of diffuse neutrino observations (per flavor) by IceCube. The black and grey data show IceCube's measurement of the atmospheric $\nu_e+\bar\nu_e$~\cite{Aartsen:2012uu,Aartsen:2015xup} and $\nu_\mu +\bar\nu_\mu$~\cite{Abbasi:2010ie} spectra.  The magenta line and magenta-shaded area indicate the best-fit and $1\sigma$ uncertainty range of a power-law fit to the six-year HESE data. Note that the HESE analysis vetoes atmospheric neutrinos and can probe astrophysical neutrinos below the atmospheric neutrino flux. The corresponding fit to the eight-year $\nu_\mu+\bar\nu_\mu$ analysis is shown in red. Figure from Ref.~\cite{Ahlers:2018fkn}.}
\label{fig:fluxes}
\end{figure}

Using the Earth as a filter, a flux of neutrinos has been identified that is predominantly of atmospheric origin. IceCube has measured this flux over three orders of magnitude in energy with a result that is consistent with theoretical calculations. However, with eight years of data, an excess of events is observed at energies beyond 100\,TeV~\cite{Aartsen:2015rwa,Aartsen:2016xlq,Aartsen:2017mau}, which cannot be accommodated by the atmospheric flux; see Fig.~\ref{fig:fluxes}. Allowing for large uncertainties on the extrapolation of the atmospheric component to higher energy, the statistical significance of the excess astrophysical flux is $6.7\sigma$. While IceCube measures only the energy of the secondary muon inside the detector, from Standard Model physics we can infer the energy spectrum of the parent neutrinos. The cosmic neutrino flux is well described by a power law with a spectral index $\Gamma=2.19\pm0.10$ and a normalization at 100\,TeV neutrino energy of $(1.01^{+0.26}_{-0.23})\,\times10^{-18}\,\rm GeV^{-1}\rm cm^{-2} \rm sr^{-1}$~\cite{Aartsen:2017mau}. The neutrino energies contributing to this power-law fit cover the range from 119\,TeV to 4.8\,PeV.

Using only two years of data, it was the alternative HESE method, which selects neutrinos interacting inside the detector, that revealed the first evidence for cosmic neutrinos~\cite{Aartsen:2013bka,Aartsen:2013jdh}. The segmentation of the detector into a surrounding veto and active signal region has been optimized to reduce the background of atmospheric muons and neutrinos to a handful of events per year, while keeping most of the cosmic signal. Neutrinos of atmospheric and cosmic origin can be separated not only  by using their well-measured energy but also on the basis that background atmospheric neutrinos reaching us from the Southern Hemisphere can be removed because they are accompanied by particles produced in the same air shower where the neutrinos originate. A sample event with a light pool of roughly one hundred thousand photoelectrons extending over more than 500 meters is shown in the left panel of Fig.~\ref{fig:erniekloppo}. With PeV energy, and no trace of accompanying muons from an atmospheric shower, these events are highly unlikely to be of atmospheric origin. The six-year data set contains a total of 82 neutrino events with deposited energies ranging from 60\,TeV to 10\,PeV. The data are consistent with an astrophysical component with a spectrum close to $E^{-2}$ above an energy of $\sim 200$\,TeV. In summary, IceCube has observed cosmic neutrinos using both methods for rejecting background. Based on different methods for reconstruction and energy measurement, their results agree, pointing at extragalactic sources whose flux has equilibrated in the three flavors after propagation over cosmic distances~\cite{Aartsen:2015ivb} with $\nu_e:\nu_\mu:\nu_\tau \sim 1:1:1$.

\begin{figure}[t]\centering
\includegraphics[width=0.95\linewidth,viewport=5 30 645 360,clip=true]{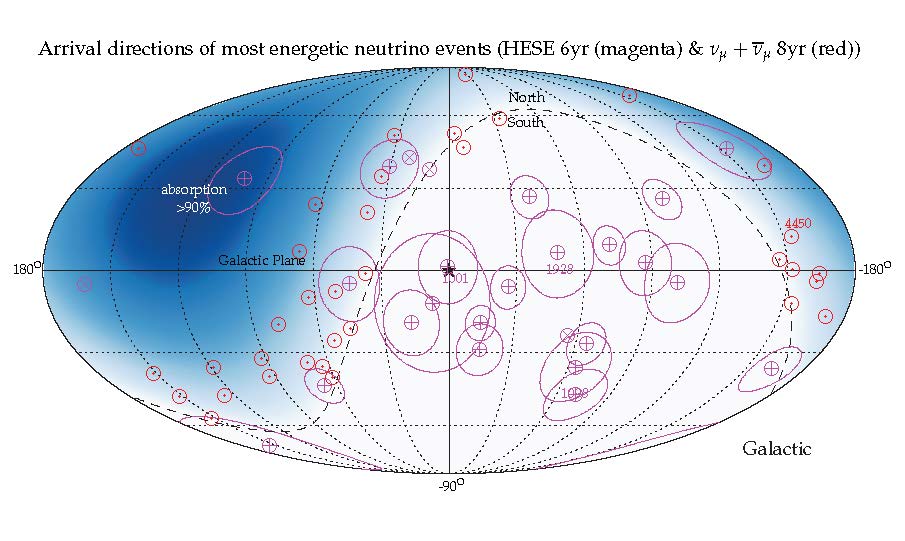}
\caption[]{Mollweide projection in Galactic coordinates of the arrival direction of neutrino events. We show the results of the eight-year upgoing track analysis~\cite{Aartsen:2017mau} with reconstructed muon energy $E_\mu\gtrsim200$~TeV ({$\odot$}). The events of the six-year high-energy starting event (HESE) analysis with deposited energy larger than 100\,TeV (tracks {$\otimes$} and cascades {$\oplus$}) are also shown~\cite{Aartsen:2014gkd,Aartsen:2015zva,Aartsen:2017mau}. The thin circles indicate the median angular resolution of the cascade events ({$\oplus$}). The blue-shaded region indicates the zenith-dependent range where Earth absorption of 100~TeV neutrinos becomes important, reaching more than 90\% close to the nadir. 
The dashed line indicates the horizon and the star ({\scriptsize $\star$}) the Galactic Center. We highlight the four most energetic events in both analyses by their deposited energy (magenta numbers) and reconstructed muon energy (red number). Figure from Ref.~\cite{Ahlers:2018fkn}.}
\label{fig:NeutrinoMap}
\end{figure}

 An extrapolation of this high-energy flux to lower energy suggests an interesting excess of events in the $30-100$\,TeV energy range over and above a single power-law fit; see Fig.~\ref{fig:NeutrinoMap}. This conclusion is supported by a subsequent analysis that has lowered the threshold of the starting-event analysis~\cite{Aartsen:2016tpb} and by a variety of other analyses. The astrophysical flux measured by IceCube is not featureless; either the spectrum of cosmic accelerators cannot be described by a single power law or a second component of cosmic neutrino sources emerges in the spectrum. Because of the self-veto of atmospheric neutrinos in the HESE analysis, i.e.,~the veto triggered by accompanying atmospheric muons, it is very difficult to accommodate the component below 100\,TeV as a feature in the atmospheric background.

In Figure \ref{fig:NeutrinoMap} we show the arrival directions of the most energetic events in the eight-year upgoing $\nu_\mu+\bar\nu_\mu$ analysis ($\odot$) and the six-year HESE data sets. The HESE data are separated into tracks ($\otimes$) and cascades ($\oplus$). The median angular resolution of the cascade events is indicated by thin circles around the best-fit position. The most energetic muons with energy $E_\mu>200$~TeV in the upgoing $\nu_\mu+\bar\nu_\mu$ data set accumulate near the horizon in the Northern Hemisphere. Elsewhere, muon neutrinos are increasingly absorbed in the Earth causing the apparent anisotropy of the events in the Northern Hemisphere. Also HESE events with deposited energy of $E_{\rm dep}>100$~TeV suffer from absorption in the Earth and are therefore mostly detected when originating in the Southern Hemisphere. After correcting for absorption, the arrival directions of cosmic neutrinos are isotropic,  suggesting extragalactic sources. In fact, no correlation of the arrival directions of the highest energy events, shown in Fig.~\ref{fig:NeutrinoMap}, with potential sources or source classes has reached the level of $3\sigma$~\cite{Aartsen:2016tpb}.

\section{IceCube neutrinos and Fermi photons}\label{IC_Fermi}

The most important message emerging from the IceCube measurements of the high-energy cosmic neutrino flux is not apparent yet: the prominent and surprisingly important role of protons relative to electrons in the nonthermal universe. Photons are produced in association with neutrinos when accelerated cosmic rays produce neutral and charged pions in interactions with target photons or nuclei in the vicinity of the accelerator. Targets include strong radiation fields that may be associated with the accelerator as well as concentrations of matter, such as molecular clouds in their vicinity. Additionally, pions can be produced in the interaction of cosmic rays with the extragalactic background light (EBL) when propagating through the interstellar or intergalactic background. As already discussed in section \ref{sec1}, a high-energy flux of neutrinos is produced in the subsequent decay of charged pions via $\pi^+\to\mu^++\nu_\mu$ followed by $\mu^+ \to e^++\nu_e+\bar\nu_\mu$ and the charge--conjugate processes. High-energy gamma rays result from the decay of neutral pions, $\pi^0\to\gamma+\gamma$. Pionic gamma rays and neutrinos carry, on average, 1/2 and 1/4 of the energy of the parent pion, respectively. With these approximations, the neutrino production rate $Q_{\nu_\alpha}$ (units of ${\rm GeV}^{-1} {\rm s}^{-1}$) can be related to the one for charged pions as
\begin{equation}\label{eq:PIONtoNU}
\sum_{\alpha}E_\nu Q_{\nu_\alpha}(E_\nu) \simeq 3\left[E_\pi Q_{\pi^\pm}(E_\pi)\right]_{E_\pi \simeq 4E_\nu}\,.
\end{equation}
Similarly, the production rate of pionic gamma-rays is related to the one for neutral pions as
\begin{equation}\label{eq:PIONtoGAMMA}
E_\gamma Q_{\gamma}(E_\gamma) \simeq 2\left[E_\pi Q_{\pi^0}(E_\pi)\right]_{E_\pi \simeq 2E_\gamma}\,.
\end{equation}

Note, that the relative production rates of pionic gamma rays and neutrinos only depend on the ratio of charged-to-neutral pions produced in cosmic-ray interactions, denoted by $K_\pi = N_{\pi^\pm}/N_{\pi^0}$. Pion production by cosmic rays in interactions with photons can proceed resonantly in the processes $p + \gamma \rightarrow \Delta^+ \rightarrow \pi^0 + p$ and $p + \gamma \rightarrow \Delta^+ \rightarrow \pi^+ + n$. These channels produce charged and neutral pions with probabilities 2/3 and 1/3, respectively. However, the additional contribution of nonresonant pion production changes this ratio to approximately 1/2 and 1/2. In contrast, cosmic rays interacting with matter, e.g.,~hydrogen in the Galactic disk, produce equal numbers of pions of all three charges: $p+p \rightarrow N_\pi\,[\,\pi^{0}+\pi^{+} +\pi^{-}]+X$, where $N_\pi$ is the pion multiplicity. From above arguments we have $K_\pi\simeq2$ for cosmic ray interactions with gas ($pp$) and $K_\pi\simeq1$ for interactions with photons ($p\gamma$) {\cite{Ahlers:2018fkn}. 

With this approximation we can combine Eqs.~(\ref{eq:PIONtoNU}) and (\ref{eq:PIONtoGAMMA}) to derive a powerful relation between the pionic gamma-ray and neutrino production rates:
\begin{equation}\label{eq:GAMMAtoNU}
\frac{1}{3}\sum_{\alpha}E^2_\nu Q_{\nu_\alpha}(E_\nu) \simeq \frac{K_\pi}{4}\left[E^2_\gamma Q_\gamma(E_\gamma)\right]_{E_\gamma = 2E_\nu}\,.
\end{equation}
The prefactor $1/4$ accounts for the energy ratio $\langle E_\nu\rangle/\langle E_\gamma\rangle\simeq 1/2$ and the two gamma rays produced in the neutral pion decay.  

Note that this relation relates pionic neutrinos and gamma rays without any reference to the cosmic ray beam; it simply reflects the fact that a $\pi^0$ produces two $\gamma$ rays for every charged pion producing a $\nu_\mu +\bar\nu_\mu$ pair, which cannot be separated by current experiments.

Before applying this relation to a cosmic accelerator, we have to take into account the fact that, unlike neutrinos, gamma rays interact with photons of the cosmic microwave background before reaching Earth. The resulting electromagnetic shower subdivides the initial photon energy, resulting in multiple photons in the GeV-TeV energy range by the time the photons reach Earth. Calculating the cascaded gamma-ray flux accompanying IceCube neutrinos is straightforward~\cite{Protheroe1993,Ahlers:2010fw}.

As an illustration, an example of $\gamma$-ray and neutrino emission is shown as blue lines in Fig.~\ref{fig:panorama} assuming that the underlying $\pi^0$ / $\pi^\pm$ production follows from cosmic-ray interactions with gas in the universe. In this way, the initial emission spectrum of $\gamma$-rays and neutrinos from pion decay is almost identical to the spectrum of cosmic rays (assumed to be a power law, $E^{-2.19}$ as is the case for the diffuse cosmic neutrino flux above and energy of 100\,TeV), after accounting for the different normalizations and energy scales. The flux of neutrinos arriving at Earth (blue dashed line) follows this initial CR emission spectrum. However, the observable flux of $\gamma$-rays (blue solid lines) is strongly attenuated above 100~GeV by interactions with extragalactic background photons \cite{Ahlers:2018fkn}.

The overall normalization of the emission is chosen in a way that the model does not exceed the isotropic $\gamma$-ray background observed by the Fermi satellite (blue data). This implies an upper limit on the neutrino flux shown as the blue dashed line. Interestingly, the neutrino data shown in Fig.~\ref{fig:panorama} saturates this limit above 100~TeV. Moreover, the HESE data that extends to lower energies is only marginally consistent with the upper bound implied by the model (blue dashed line). This example shows that multimessenger studies of $\gamma$-ray and neutrino data are powerful tools to study the neutrino production mechanism and to constrain neutrino source models~\cite{Murase:2013rfa}.

The matching energy densities of the extragalactic gamma-ray flux detected by Fermi and the high-energy neutrino flux measured by IceCube suggest that, rather than detecting some exotic sources, it is more likely that IceCube to a large extent observes the same universe conventional astronomy does. Clearly, an extreme universe modeled exclusively on the basis of electromagnetic processes is no longer realistic. The finding implies that a large fraction, possibly most, of the energy in the nonthermal universe originates in hadronic processes, indicating a larger role than previously thought. The high intensity of the neutrino flux below 100~TeV in comparison to the Fermi data might indicate that these sources are even more efficient neutrino than gamma-ray sources~\cite{Murase:2015xka,Bechtol:2015uqb}.

Interestingly, the common energy density of photons and neutrinos is also comparable to that of the ultra-high-energy extragalactic cosmic rays (above $10^{9}$~GeV) observed, for instance, by the Auger observatory~\cite{Aab:2015bza} (green data). Unless accidental, this indicates a common origin of the signal and illustrates the potential of multimessenger studies. A scenario where the high-energy neutrinos observed at IceCube could actually originate in the same sources could be realized as follows: the cosmic ray sources can be embedded in environments that act as ``storage rooms'' for cosmic rays with energies far below the ``ankle'' ($E_{\rm CR}\ll1$EeV). This energy-dependent trapping can be achieved via cosmic ray diffusion in magnetic fields. While these cosmic rays are trapped, they can produce $\gamma$-rays and neutrinos via collisions with gas. If the conditions are right, this mechanism can be so efficient that the total energy stored in cosmic rays below the ankle is converted to that of $\gamma$-rays and neutrinos. These ``calorimetric'' conditions can be achieved, for instance, in starburst galaxies~\cite{Loeb:2006tw} or galaxy clusters~\cite{Berezinsky:1996wx}.
\begin{figure}[t]
\centering
\includegraphics[width=0.95\linewidth]{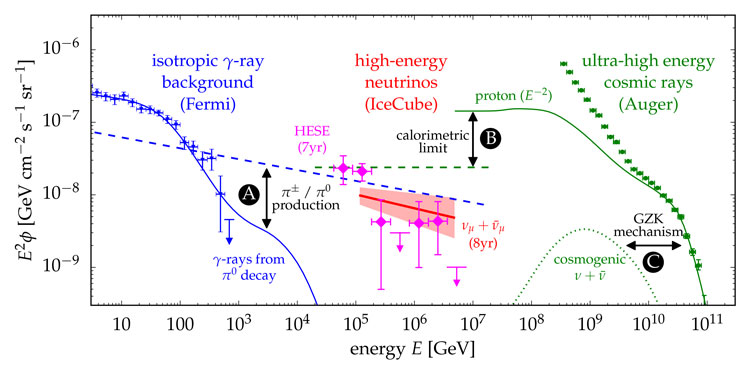}
\caption{The spectral flux ($\phi$) of neutrinos inferred from the eight-year upgoing track analysis (red fit) and the seven-year HESE analysis (magenta fit) compared to the flux of unresolved extragalactic $\gamma$-ray sources~\cite{Ackermann:2014usa} (blue data) and ultra-high-energy cosmic rays~\cite{Aab:2015bza} (green data). The neutrino spectra are indicated by the best-fit power-law (solid line) and $1\sigma$ uncertainty range (shaded range). We highlight the various multimessenger interfaces: {\bf A:} The joined {production of charged pions ($\pi^\pm$) and neutral pions ($\pi^0$)} in cosmic-ray interactions leads to the emission of neutrinos (dashed blue) and $\gamma$-rays (solid blue), respectively. {\bf B:} Cosmic ray emission models (solid green) of the most energetic cosmic rays imply a maximal flux ({calorimetric limit}) of neutrinos from the same sources (green dashed). {\bf C:} The same cosmic ray model predicts the emission of cosmogenic neutrinos from the collision with cosmic background photons ({GZK mechanism}). Figure from Ref.~\cite{Ahlers:2018fkn}.}
\label{fig:panorama}
\end{figure}

The extragalactic $\gamma$-ray background observed by Fermi~\cite{Ackermann:2014usa} has contributions from  identified point-like sources on top of an isotropic $\gamma$-ray background (IGRB) shown in Fig.~\ref{fig:panorama}. This IGRB is expected to consist mostly of emission from the same class of $\gamma$-ray sources that are individually below Fermi's point-source detection threshold (see, e.g., Ref.~\cite{DiMauro:2015tfa}). A significant contribution of $\gamma$-rays associated with IceCube's neutrino observation would have the somewhat surprising implication that indeed many extragalactic $\gamma$-ray sources are also neutrino emitters, while none had been detected so far. This dramatically changed when IceCube developed methods for performing real-time multiwavelength observations in cooperation with some twenty other observatories to identify the sources and build on the discovery of cosmic neutrinos to launch a new era in astronomy~\cite{Aartsen:2016qbu,Aartsen:2016lmt}. This effort led to the identification of a distant flaring blazar as a cosmic ray accelerator in a multimessenger campaign launched by a 290\,TeV energy neutrino detected from the constellation of Orion on September 22, 2017~\cite{IceCube:2018dnn}.

\section{The first truly multimessenger campaign}

Neutrinos only originate in environments where protons are accelerated to produce pions and other particles that decay into neutrinos. Neutrinos can thus exclusively pinpoint cosmic ray accelerators, and this is exactly what one neutrino did on September 22, 2017.

IceCube detects muon neutrinos, a type of neutrino that leaves a well-reconstructed track in the detector roughly every five minutes. Most of them are low-energy neutrinos produced in the Earth's atmosphere, which are of interest for studying the neutrinos themselves, but are a persistent background when doing neutrino astronomy. In 2016, IceCube installed an online filter that selects from this sample, in real time, very high energy neutrinos that are likely to be of cosmic origin \cite{Aartsen:2016lmt}. We reconstruct their energy and celestial coordinates, typically in less than one minute, and distribute the information automatically via the Gamma-ray Coordinate Network to a group of telescopes around the globe and in space for follow-up observations. These telescopes look for electromagnetic radiation from the arrival direction of the neutrino, searching for coincident emission that can reveal its origin.  

The tenth such alert~\cite{2017GCN.21916....1K}, IceCube-170922A, on September 22, 2017, reported a well-reconstructed muon neutrino with an energy of 290 TeV and, therefore, with a high probability of originating in an astronomical source. The Fermi telescope detected a flaring blazar aligned with the cosmic neutrino within 0.06 degrees. The source is a known blazar, a supermassive black hole spitting out high-energy particles in twin jets aligned with its rotation axis which is directed at Earth. This blazar, TXS 0506+056, had been relatively poorly studied until now, although it was identified as the highest energy gamma ray source detected by EGRET from any blazar with two photons above 40 GeV~\cite{Dingus:2001hz}.  The set of observations triggered by the September 22 neutrino has yielded a treasure trove of multiwavelength data that will allow us to probe the physics of the first cosmic ray accelerator. An optical telescope eventually measured its distance~\cite{Paiano:2018qeq}, which was found to be 4 billion light-years. Its large distance points to a special galaxy, which sets it apart from the ten-times-closer blazars such as the Markarian sources that dominate the extreme gamma-ray sky observed by NASA's Fermi satellite.

TXS 0506+056 was originally flagged by the Fermi~\cite{2017ATel10791....1T} and Swift~\cite{2017ATel10792....1E} satellite telescopes. Follow-up observations with the MAGIC air Cherenkov telescope~\cite{2017ATel10817....1M} identified it as a rare TeV blazar with the potential to produce the very high energy neutrino detected by IceCube. The source was subsequently scrutinized in X-ray, optical, and radio wavelengths. This is a first, truly multimessenger observation: none of the instruments could have made this breakthrough independently. In total, more than 20 telescopes observed the flaring blazar as a highly variable source in a high state~\cite{IceCube:2018dnn}.

It is important to realize that nearby blazars like the Markarian sources are at a redshift that is ten times smaller, and therefore TXS 0506+056, with a similar flux despite the greater distance, is one of the most luminous sources in the Universe. It likely belongs to a special class of blazars that accelerate proton beams as revealed by the neutrino. This is further supported by the fact that a variety of previous attempts to associate the arrival directions of cosmic neutrinos with a variety of Fermi blazar catalogues failed.

Informed by the multimessenger campaign, IceCube searched its archival neutrino data up to and including October 2017 in the direction of IC170922 using the likelihood routinely used in previous searches. This revealed a spectacular burst of over a dozen high-energy neutrinos in 110 days in the 2014-2015 data with a spectral index consistent with the one observed for the diffuse cosmic neutrino spectrum \cite{IceCube:2018cha}.

Interestingly, the AGILE collaboration, which operates an orbiting X-ray and gamma ray telescope, reported a day-long flare in the direction of a previous neutrino alert sent on Juli 31, 2016 \cite{Lucarelli:2017hhh}. The flare occurred more than one day before the time of the alert. In light of the rapid daily variations observed near the peak emission of the TXS 0506+056 flare at the time of IC170922A, this may well be a genuine coincidence.

\section{Flaring sources and the high-energy neutrino flux}
The extraordinary detection of more than a dozen cosmic neutrinos in the 2014 flare despite its 0.34 redshift further suggests that TXS 0506+056 belongs to a special class of sources that produce cosmic rays. The single neutrino flare dominates the flux of the source over the 9.5 years of archival IceCube data, leaving IC170922 as a less luminous second flare in the sample. We will show that a subset of about 5\% of all blazars bursting once in 10 years at the level of TXS 0506+056 in 2014, can accommodate the diffuse flux cosmic neutrino flux observed by IceCube. We already pointed out in the previous section that the energy density corresponding to this flux is similar to the one in the highest energy cosmic rays. 

In order to calculate the flux of high-energy neutrinos from a population of sources, we follow \cite{Halzen2002} and relate the diffuse neutrino flux to the injection rate of cosmic rays and their efficiency to produce neutrinos in the source. For a class of sources with density $\rho$ and neutrino luminosity $L_\nu$, the all-sky neutrino flux is
\begin{eqnarray}
\sum_{\alpha} E_\nu^2 \frac{d N_\nu}{dE_\nu} = \frac{1}{4\pi} \frac{c}{H_0} \xi_z L_\nu \rho,
\end{eqnarray}
where $\xi_z$ is a factor of order unity that parametrizes the integration over the redshift evolution of the sources. 
 The relation can be adapted to a fraction $\mathcal{F}$ of sources with episodic emission of flares of duration $\Delta t$ over a total observation time $T$:
\begin{eqnarray}
\sum_{\alpha} E_\nu^2 \frac{d N_\nu}{dE_\nu} = \frac{1}{4\pi} \frac{c}{H_0} \xi_z L_\nu \rho \mathcal{F} \frac{\Delta t}{T} \,.
\end{eqnarray}
Applying this relation to the 2014 TXS 0506+056 burst that dominates the flux over the 9.5 years of neutrino observations, yields
\begin{eqnarray}\label{diffuse_flux}
\begin{aligned}
3\times10^{-11}\, {\rm{TeV cm^{-2} s^{-1} sr^{-1}}} = &\frac{ \mathcal{F}}{4\pi} \bigg(\frac{R_H}{3\, \rm{Gpc}} \bigg) \bigg(\frac{\xi_z}{0.7 \rm{}} \bigg) \bigg(\frac{L_\nu}{1.2\times10^{47}\, \rm{erg/s}} \bigg)\\ &\bigg(\frac{\rho}{1.5\times10^{-8}\, \rm{Mpc^{-3}}} \bigg) \bigg(\frac{\Delta t}{110 \,{\rm d}} \frac{10 {\, \rm yr}}{T}\bigg)\,, 
\end{aligned}
\end{eqnarray}
a relation which is satisfied for $\mathcal{F}\sim0.05$. In summary, a special class of BL Lac blazars that undergo $\sim110$-day duration flares like TXS 0506+056 once every 10 years accommodates the observed diffuse flux of high-energy cosmic neutrinos. The class of such neutrino-flaring sources represents 5\% of the sources. The argument implies the observation of roughly 100 muon neutrinos per year. This is exactly the flux of cosmic neutrinos that corresponds to the $E^{-2.19}$ diffuse flux measured above 100\,TeV. (Note that the majority of these neutrinos cannot be separated from the atmospheric background, leaving us with the reduced number of very high energy events discussed in the previous sections).

As previously discussed, he energetics of the cosmic neutrinos is matched by the energy of the highest energy cosmic rays, with ~\cite{HKW}
\begin{eqnarray}
\frac{1}{3}\sum_{\alpha} E_\nu^2 \frac{d N_\nu}{dE_\nu} \simeq \frac{c}{4 \pi}\,\bigg( \frac{1}{2}
(1-e^{-f_\pi})\, \xi_z t_H \frac{dE}{dt} \bigg)\,.
\end{eqnarray}
The cosmic rays' injection rate $dE/dt$ above $10^{16}$ eV is $(1-2) \times 10^{44}\,\rm erg$\, $\rm Mpc^{-3}\,yr^{-1}$ \cite{Ahlers:2012rz, Katz:2013ooa}. From Eq.~\ref{diffuse_flux} it follows that the energy densities match for a pion production efficiency of the neutrino source of $f_\pi \gtrsim 0.4$. This high efficiency requirement is consistent with the premise that a special class of efficient sources is responsible for producing the high-energy cosmic neutrino flux seen by IceCube. The sources must contain sufficient target density in photons, possibly protons, to generate the large value of $f_\pi$. It is clear that the emission of flares producing the large number of cosmic neutrinos detected in the 2014 burst must correspond to major accretion events onto the black hole lasting a few months. The pionic photons will lose energy in the source and the neutrino emission is not accompanied by a flare as was the case for the 2017 event; the Fermi data, consistent with the scenario proposed, reveal photons with energies of tens of GeV, but no flaring activity.

A key question is whether the neutrino and gamma ray spectra for the 2014 neutrino burst from TXS 0506+056 satisfy the multimessenger relationship introduced in section \ref{IC_Fermi}. With the low statistics of the very high-energy gamma ray measurements during the burst period, the energetics is a more robust measure for evaluating the connection, especially because the source is opaque to high-energy gamma rays, as indicated by the large value of  $f_\pi$, and the pionic gamma rays will lose energy inside the source before cascading in the microwave photon background; for details see Ref.~\cite{HKW}. 

It is worth noting that this model for the diffuse neutrino flux clarifies why earlier attempts to associate it with blazars were unsuccessful; see Ref.~\cite{Kowalski:2014zda, Mertsch:2016hcd}. Clearly, the time-integrated studies are not applicable to time-dependent sources. Moreover, with a subclass of more energetic sources with lower density responsible for the diffuse flux, the constraints on blazars obtained from the relation between the point source limits and the diffuse flux are mitigated. Additional issues with this limit arise from the use of point source sensitivity (defined as 90\% C.L.) rather than discovery potential (defined as 5$\sigma$ C.L.) as a constraint. The former does not provide a robust statistical measure. It has been argued that the limits from the nonobservation of a point source in time-integrated searches would suppress the contribution of blazars to the total neutrino emission because the time-averaged flux of sources is directly correlated with bursts \cite{Murase:2018iyl}. However, two points are missed in this argument. First, this argument only applies if the duty cycle of the source and the duration of the flare, supplies neutrino bursts shorter than the total time that the detector has been taking data. Second, it neglects the excess buried under the atmospheric background which might not be statistically compelling yet. For instance, in the case of TXS 0506+056, the time-integrated search reports a significance of 2.1$\sigma$ only.

We are aware that these speculations are qualitative and that they may be premature and the hope is that multimessenger astronomy will provide us with more clues after the breakthrough event of September 22, 2017 that generated an unmatched data sample over all wavelengths of the electromagnetic spectrum on the first identified cosmic ray accelerator.

\section{Summary}
Getting all the elements of this puzzle to fit together is not easy, but they suggest that the blazar may contain important clues on the origin of cosmic neutrinos and cosmic rays. This breakthrough is just the beginning and raises intriguing questions. What is special about this source? Can a subclass of blazars to which it belongs accommodate the diffuse flux observed by IceCube? Are these also the sources of all high-energy cosmic rays or only of some? The TXS 0506+056 emission over the last 10 years is dominated by the single flare in 2014. If this is characteristic of the subclass of sources that it belongs to, identifying additional sources will be difficult unless more and larger neutrino telescopes  yield more frequent and higher statistics neutrino alerts. 

Unlike the previous SN1987A and GW170817 multimessenger events, this event could not have been observed with a single instrument. Without the initial coincident observation, IC170922A would be just one more of the few hundred cosmic neutrinos detected by IceCube and the accompanying radiation just one more flaring blazar observed by Fermi-LAT. Neutrino astronomy was born with a supernova in 1987. Thirty years later, this recent event involves neutrinos that are tens of millions of times more energetic and are from a source a hundred thousand times more distant.

\bibliographystyle{style}
\bibliography{bib}

\end{document}